# Multi-tree Quantum Routing in Realistic Topologies

Zebo Yang, *Graduate Student Member, IEEE*, Ali Ghubaish, *Graduate Student Member, IEEE*, and Raj Jain, *Life Fellow, IEEE*, Ramana Kompella, *Member, IEEE,* and Hassan Shapourian, *Member, IEEE*

*Abstract*—In entanglement distribution networks, communication between two nodes necessitates the generation of end-to-end entanglement by entanglement swapping at intermediate nodes. Efficiently creating end-to-end entanglements over long distances is a key objective. In our prior study on asynchronous routing, we enhanced these entanglement rates by leveraging solely the local knowledge of the node's entanglement links. This was achieved by creating a tree structure, particularly a destination-oriented directed acyclic graph (DODAG) or a spanning tree, eliminating synchronous operations and conserving unused entanglement links. In this article, we present a multi-tree approach with multiple DODAGs designed to improve end-to-end entanglement rates in large-scale networks, specifically catering to a range of network topologies, including grids and barbells, as well as realistic topologies found in research testbeds like ESnet and Internet2. Our simulations show a marked improvement in end-to-end entanglement rates for specific topologies compared to the single-tree method. This study underscores the promise of asynchronous routing schemes in quantum networks, highlighting the effectiveness of asynchronous routing across different network topologies and proposing a superior routing tactic.

## I. Introduction

ADVANCEMENTS in quantum technology have ushered in new horizons for communication networks [1], [2]. Quantum networks, underpinned by the principle of quantum entanglement, present the potential for unparalleled security and computational advantages over classical counterparts. Pioneering applications, such as quantum key distribution (QKD) [3] and distributed quantum computations [4], are a testament to these capabilities. Essential to these applications is creating end-to-end entanglement over long distances, also known as entanglement distribution. This challenge is being addressed by the development of quantum repeaters [5], [6]. These repeaters employ entanglement swapping [7], enabling two quantum nodes that are not directly connected to establish entanglement via an intermediate node, i.e., the repeater. This procedure can be extended over multiple repeaters, creating a long-distance entanglement link.

In prevalent methods [8], [9], the routing procedures bifurcate into two distinct phases: the external and the internal, each with a designated time slot. In the external phase, node pairs with entangled photons create a direct link between them. This results in an *instant topology* of these successfully established direct-link entanglements. Fig. 1a shows an initial state (*physical topology*) which results in several direct links, e.g., $E_{AD}$ shown in Fig. 1b. Transitioning to the internal phase, under the presumption that nodes are only aware of the generation outcomes of the direct-link entanglements with their adjacent nodes (referred to as local knowledge of the instant topology), each repeater node indiscriminately swaps entanglements, aiming to establish a multi-hop entanglement between two communication-requesting endpoints as illustrated in Fig. 1c and 1d. $E_{AC}$ in Fig. 1d is an example. If the nodes A and C were wanting to communicate, this would be the desired end-to-end entanglement. Should this end-to-end entanglement fall short, the two phases would repeat. To maintain the availability of entanglement links for the internal phase, synchronizing both phases becomes imperative. Yet, such synchronization and indiscriminate trials expend all available entanglements within each timeframe, resulting in a diminished end-to-end entanglement rate.

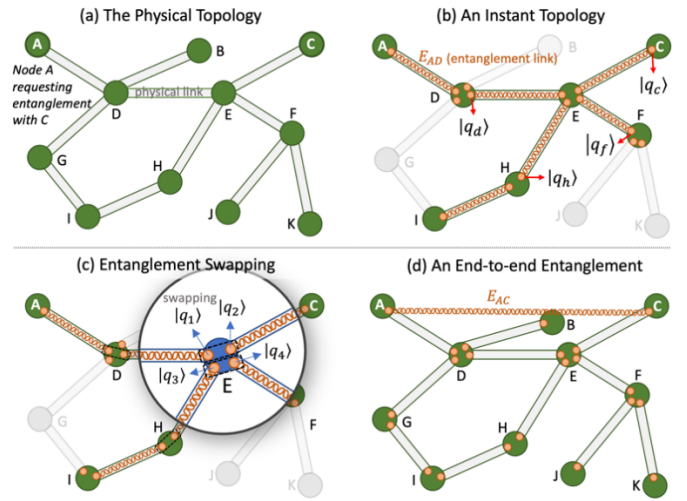

**Fig. 1.** Example of end-to-end entanglement generation.

In [10], we proposed an asynchronous routing protocol that identifies paths between any two endpoints at a substantially enhanced entanglement rate compared to the synchronized two-phase methods. Additionally, as quantum technology advances, our rate is expected to increase along with extended coherence times. Moreover, the incorporation of root nodes as gateways can facilitate the integration of multiple interconnected networks. By iteratively employing an instant topology built on entanglement links, our approach obviates the need for synchronized operations. This instant topology can manifest as a tree-like graph akin to a destination-oriented directed acyclic graph (DODAG) [10] or a spanning tree. Our theoretical evaluations and simulations showed that asynchronous methods outpace their synchronous counterparts by achieving a larger upper bound and an increased end-to-end entanglement rate.

*Corresponding author: Zebo Yang
Zebo Yang, Ali Ghubaish, and Raj Jain are with the Department of Computer Science and Engineering, Washington University in St. Louis, St. Louis, MO, USA (e-mail: zebo@wustl.edu; aghubaish@wustl.edu; jain@wustl.edu)
Ramana Kompella and Hassan Shapourian are with Cisco Research, San Jose, CA, USA (e-mail: rkompell@cisco.com; hshapour@cisco.com).

While the asynchronous routing scheme has introduced a practical approach using a single tree-like structure, relying on a single DODAG may introduce limitations, especially regarding scalability and resilience in an expanding quantum network landscape. As networks grow in complexity and size, there arises a need for multiple interconnected trees to achieve better routing efficiency, fault tolerance, and quicker end-to-end entanglement generation. In this paper, we propose the integration of multiple trees (multiple DODAGs) for quantum network routing and highlight its advantages over the single-tree approach and synchronous methods. Our simulations indicate a significant enhancement in end-to-end entanglement rates for specific topologies, including grid, barbell, and some realistic testbed structures using the multi-tree approach. We also identify the optimal graph types for tree routing schemes, which feature evenly distributed nodes with more uniform and higher connectivity. This research emphasizes the potential of asynchronous routing methods in quantum networks, spotlighting the success of asynchronous routing in various network structures and advocating advanced routing strategies.

Note that, in this paper, when referring to the entanglement rate, we are addressing the end-to-end entanglement rate across multiple hops rather than the direct entanglement generation rate (physical-layer-wise) between two nodes.

## II. BACKGROUND

The success of quantum networks heavily hinges upon the ability to transmit quantum information across long distances. The storage and transmission of quantum bits (qubits) suffer from many issues [2], the most prominent being decoherence and photon loss. Quantum repeaters are to overcome these challenges. In this section, we provide the research context.

### A. Entanglement Swapping by Repeaters

Entanglement swapping is a quantum operation that enables two particles to become entangled without direct interaction. By employing an intermediary quantum repeater that is entangled with both particles, a measurement on the repeater can result in the two particles becoming entangled. As illustrated in Fig. 1b, several entanglement links are present, like $E_{AD}$. The goal is to convert these "direct links" into multi-hop entanglements, connecting farther nodes and establishing an end-to-end entanglement between the communication-requesting nodes, in our case, Nodes A and C. For instance, we might aim to establish entanglement between Nodes D and C via Node E, leveraging the entanglement links $E_{DE}$ and $E_{EC}$. In this scenario, Node E would execute an entanglement swapping operation on its qubits $|q_1\rangle$ and $|q_2\rangle$. If successful, $E_{DE}$ and $E_{EC}$ are consumed, and a new entanglement $E_{DC}$ is created. Also, Node E would conduct another entanglement swapping to produce the desired end-to-end entanglement, $E_{AC}$.

These repeater nodes embody a fusion of quantum and classical systems, where the classical systems handle routing. They create end-to-end entanglements for user pairs. Then, the end users use the end-to-end entanglement as a communication resource. For simplicity, we assume every network node is a repeater node in this study.

### B. The Routing Problem for Quantum Networks

Consider a graph $G(V, E)$ symbolizing the physical layout of the quantum network, as shown in Fig. 1a. Here, every node $v \in V$ is a repeater node, and each edge $e \in E$ signifies a physical channel between two adjacent repeaters. The instant topology is labeled as $G'(V', E')$ and is a subset of $G$. The set $E'$ comprises direct-link entanglements, while $V'$ includes nodes linked by these entanglements. Notably, each edge $e$ can facilitate entanglement generation by a qubit pair between neighboring nodes. Every node has a limited set of qubits, represented as orange circles in Fig. 1. We simplify the problem by assuming a single qubit on each side of a physical link, as shown in Fig. 1. Also, every entanglement has an average coherence time, $T_{co}$, which refers to the duration an entanglement can persist without significant decay or loss due to environmental disturbances. Moreover, we focus on single connection requests for simplicity, leaving the study of multiple simultaneous requests for future work.

Each entanglement generation on a direct link $e$ has a success probability $p(e)$, and every entanglement swap at repeater $v$ holds a success chance of $q(v)$. These probabilities generally depend on factors such as node distance and transmissivity. For simplicity, we assume a uniform entanglement generation probability, $p$, and an entanglement swapping probability, $q$, throughout the network. On the other hand, the end-to-end entanglement rate, symbolized by $\xi$, measures the count of end-to-end entanglements created in a time unit $T$, which should be $T \leq T_{co}$. Typically, the time needed for internal and external phases (as outlined earlier) is considered a unit time, which is assumed to be less than or equal to the coherence time. To ease comparisons, we define the unit time as $T = T_{co}/n$, with integer $n \in \mathbb{Z}$ and $n \geq 1$. Here, $n$ is an integer representing the duration of the coherence time. For instance, in simulations, if $T_{co} = n$, it indicates that the coherence time for that specific simulation spans $n$ unit time. Note that $\xi$ is for the entanglement between arbitrary nodes, typically the source and destination, regardless of whether they share a direct quantum channel. Maximizing $\xi$ is our goal to enhance the likelihood of user pairs being connected by an end-to-end entanglement.

With that, quantum routing aims to find a way to optimize the end-to-end entanglement rate. Existing protocols, such as [8] and [9], rely on the global knowledge of the instant topology, necessitating that every node be aware of the complete direct-link generation status before routing (essentially, a holistic view of the instant topology). As the network's size increases, this approach requires entanglements with extended coherence times due to the lengthy process of disseminating direct-link entanglement information throughout the network. In expansive quantum networks, by the time the full direct-link generation results have spread, the direct-link entanglements may have already deteriorated.

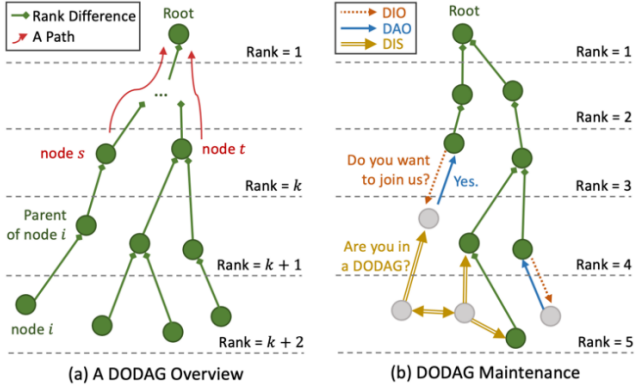

**Fig. 2.** DODAG and its maintenance.

Our prior work [10], introduced in Section II.C, suggests an asynchronous routing strategy, with nodes only informed about adjacent links in the current instant topology. For instance, Node E in Fig. 1b is aware only of the direct-link generation outcomes of $E_{ED}$, $E_{EH}$, $E_{EF}$, and $E_{EC}$. A method in [8] also takes into account merely the local knowledge of the instant topology. Still, it demands synchronized actions and depletes all redundant direct-link entanglements, resulting in a reduced rate compared to our asynchronous method.

*C. Asynchronous Routing for Quantum Networks*

This subsection outlines the key aspects of our previous work in [10] necessary for comprehending this approach. At its core, our asynchronous routing approach constructs and upholds the instant topology as a tree graph (examples include DODAG or spanning tree) and responds to connection requests by adhering to the route outlined in the graph, such as navigating through the latest DODAG's root. The critical aspect of this method is the initial service, detailing how to sustain the instant topology in the form of a distributed graph that delineates the route for any node pair. As the instant topology undergoes regular updates based on direct-link entanglement generation, a source node initiates a connection request to set up end-to-end entanglement with a destination node. Each intermediate node then instructs the next node in the path to the destination to carry out an entanglement swapping, and this process persists until the destination is reached. When a link in instant topology decoheres, the disconnected end node will be removed from the tree. If the removed node is still connected to some nodes, the branch will remain as such until one end in that branch reconnects to the tree. The branch's rank will then be updated based on the new rank of the reconnecting end node.

A DODAG is a tree-structured graph initiated from a root node and expanded through the exchange of control messages, depicted in Fig. 2. This routing approach, also used by Routing over Low-Power Lossy-Links (RPL) working group of Internet Engineering Task Force (IETF), hinges on using one or several DODAGs, leveraging specific *rank values* to help choose the best route to the root node. Communication between two endpoints is then facilitated through this root node.

A node's rank in a DODAG indicates its connectivity with the root and is determined as a scalar measure. As the node and link qualities are considered homogenous in this paper, the ranks should increase monotonically as the nodes get farther from the root, as shown in Fig. 2. It is worth noting that a DODAG node can have multiple parents if there are no loops in the graph, as is discussed in the appendix in [10]. Thus, routing can select preferred paths based on factors such as fewer hops or stronger nodes or links when dealing with distinct node or link qualities.

Fig. 2a illustrates a DODAG instance at a specific moment, showing node ranks increasing with distance from the root node. Green nodes represent nodes already incorporated into the tree, while grey nodes await inclusion. Nodes within the tree can serve as parent nodes to others, e.g., grey nodes in Fig. 2b. The grey nodes engage with adjacent nodes by dispatching DIS (DODAG Information Solicitation) messages to determine their inclusion in the DODAG. Upon receiving confirmation via DIO (DODAG Information Object) messages from a neighbor, meaning it is in the DODAG, the inquiring node sends DAO (Destination Advertisement Object) messages to request its inclusion in the DODAG. The neighbor then assigns a rank to the aspiring member and grants entry. In cases of unresponsive neighbors, the node persists in sending DIS messages until it receives a positive response or is invited into the DODAG by another DODAG neighbor through DIO.

As the DODAG undergoes continuous updates, the routing procedure remains straightforward: progress toward the root before moving to the intended endpoint, selecting the parent with the most minimal rank. This is illustrated by the red paths from Node $s$ to Node $t$ via the root node in Fig. 2a.

III. MULTI-TREE ENTANGLEMENT ROUTING

This section explores integrating a multi-tree structure with DODAG in our asynchronous routing framework, addressing the root node selection and the multi-tree applicability.

*A. Multi-tree Structure*

A multi-tree structure comprises several trees that may share subtrees while avoiding the creation of loops. When combined with DODAG, it forms a collection of DODAGs with nodes organized hierarchically based on their rank values, as illustrated in Fig. 3a. Each DODAG generates its own tree, originating from its root node. When two leaf nodes intersect, they engage in a negotiation to determine which one will become the parent, typically by comparing their rank values. The leaf node with the smaller rank value is selected as the parent. Nodes with identical rank values are called "comparable nodes" and do not establish connections with each other. The subtrees that emerge after the patterned purple nodes in Fig. 3a represent the shared subtrees.

The hierarchical node arrangement by rank value eliminates the possibility of loop formation inside one tree. Also, diamond patterns are prohibited to avoid loops between trees. For example, the subgraph outlined by the dotted ellipse in Fig. 3a demonstrates this prohibition. A diamond pattern occurs when a node with parents from multiple trees has more than one parent within a single tree. This cannot be allowed because when a source node seeks a destination path, it explores both lower and higher-ranked directions, creating an undirected crosstree path. This undirected path would lead to loop formation during the pathfinding phase [11]. To prevent these patterns, nodes refrain from being overly selective when choosing parents. In other words, a node should only select one parent per tree. For example, in Fig. 3a, if Node E already has a parent in the fourth tree, it should abstain from creating another entanglement in the tree, such as $E_{DE}$, with Node D.

In this context, finding a path between nodes entails identifying a shared parent, with a preference for selecting lower-ranked parents. The search extends to higher-ranked nodes if no common parent is located among the lower-ranked nodes. For instance, consider Node A in Fig. 3a, aiming to establish entanglement with Node B. The path for end-to-end entanglement is indicated by the yellow dashed arrows. Initially, it attempts to locate Node B among its lower-ranked peers but fails. Then, Node A extends the search to higher-ranked nodes, seeking a child node with an outgoing link to another tree. Node A finally identifies Node G and routes through it to reach Node C, which serves as a common parent for both Nodes A and B. This process is indicated by the dotted yellow arrows, signifying the undirected path from Node A to Node C and from Node B to Node C. It is worth noting that this pathfinding operates using classical methods. Once the path is established, entanglement swapping along the path is executed to create the end-to-end entanglement.

It is worth noting that the path search stops upon reaching the destination in most cases, but in the worst-case scenario, the source node may explore the entire forest. Even then, it only provides multiple paths for selection, not attempting swapping throughout. The source can choose the shortest one and perform swapping accordingly. The exponential decay of performance occurs during swapping, not during classical communication. However, in the worst-case scenario, most methods, including ours, may encounter issues with classical communication overhead in large networks, where no perfect solution exists. In the noisy intermediate-scale quantum (NISQ) era, we focus on optimizing performance despite these noisy conditions.

The first topology usually considered in routing is that of a grid. In a typical grid topology, nodes are neatly surrounded by neighboring nodes, typically four: north, south, east, and west. With that, the initial step involves the selection of root nodes, ideally one per quadrant in larger grids, while a central root may suffice for smaller grids. These roots serve as primary sources of information flow. Each layer of trees extends outward from these roots, including nodes at increasing distances. Due to the grid's symmetry, this arrangement results in concentric trees of nodes around each root. Fig. 3b shows an example of four trees in a grid topology corresponding to the one depicted in Fig. 3a.

The inherent predictability of grid layouts and the structured deployment of DODAGs minimizes the occurrence of lengthy, convoluted routes, especially for nodes situated far from one root but closer to another root within a multi-tree structure. Consequently, shorter routes reduce probabilistic operations, such as entanglement swapping, which, in turn, contributes to an increased entanglement rate.

However, employing multiple DODAGs within a grid also presents challenges. The overlapping zones of neighboring DODAGs can lead to communication overheads as nodes actively negotiate parent selection. This issue can be mitigated by discouraging the greedy selection of parents in different DODAGs or using priority and scheduling mechanisms.

*B. Root Nodes Selection*

The root node selection is crucial in optimizing entanglement distribution and minimizing overhead in any network topology, including multiple or single-tree schemes. For example, central nodes often have equidistant access to all other nodes in the grid, ensuring an even distribution of resources and minimizing the maximum path length to any node. Corner nodes allow for a clear directional flow of information across the grid.

Regarding a multi-tree structure, selecting roots in denser areas of the network is preferable. These areas, characterized by higher interaction rates, allow for more efficient entanglement swapping due to the closer proximity of nodes. Clustering the network by node density and designating the centers of these clusters as root nodes can significantly improve the network's entanglement distribution. These central cluster nodes serve as effective hubs, managing local traffic efficiently and reducing the need for failure-prone or long-distance connections.

With that, considering the eccentricity of nodes is another pivotal factor in the root node selection process. Eccentricity is the maximum distance between a node and any other node in the network. Nodes with low eccentricity are preferable as root nodes because they are more centrally located, thereby reducing the maximum number of entanglement swapping must occur. Moreover, nodes with high connectivity are excellent candidates for root status. Their multiple adjacent edges imply an ability to handle numerous entanglement requests and to support a high volume of entanglement-swapping processes. A high-connectivity node as a root also adds robustness to the network, offering multiple alternative pathways for entanglement distribution in case of individual link failures.

Dynamic root selection strategies can also benefit the network by adapting to evolving conditions such as traffic patterns or connectivity changes. Nodes that become central to the network's traffic flow areas can be dynamically chosen as root nodes to optimize distribution routes (if the instant topology allows). Additionally, periodic rotation of root nodes can mitigate the risks of network bottlenecks and single points of failure. It also distributes operational stress evenly across the network, which is especially important if the role of a root node causes any form of degradation or wear over time.

*C. Applicability*

The use of multi-tree structures offers advantages in routing entanglement. However, the specific network topology influences the choice to employ multiple trees. While multi-tree structures can be applied to various network topologies, some network configurations benefit more from this approach.

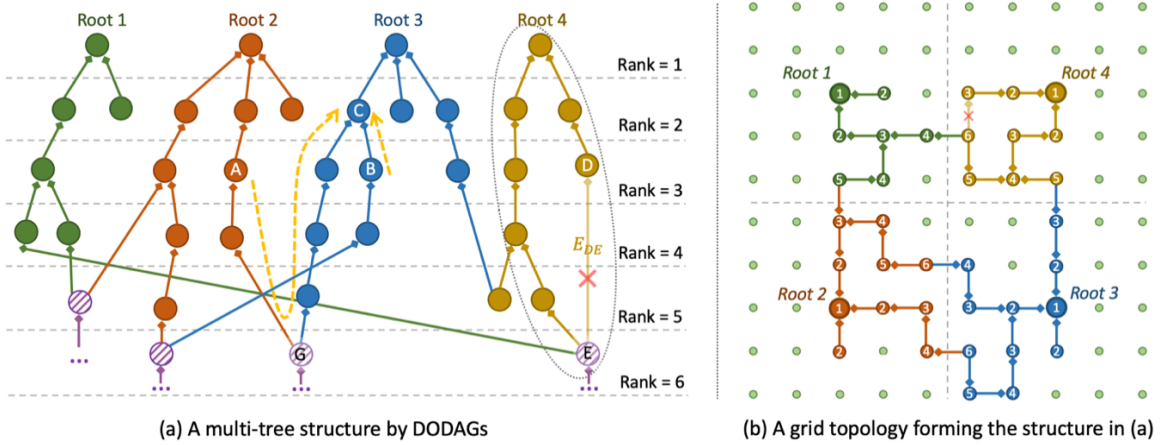

**Fig. 3.** The multi-tree structure example: a DODAG forest instance and its representation in a grid topology.

Multi-tree routing is generally more effective in larger, more regularly distributed, or densely connected topologies than linear chains or backbones. Its goal is to facilitate nodes in proximity within a large network without routing through a central root. It enhances entanglement rates for nearby nodes while maintaining rates for farther nodes. These findings are supported by the simulations presented in the next section.

Grid topologies, characterized by even connectivity, are well-suited for multi-tree routing. Using multiple trees in such grids ensures balanced load distribution. In larger grid networks, adopting a multi-tree structure significantly reduces the path length needed for end-to-end entanglement.

Conversely, multi-tree routing may not be the most competitive in barbell topologies, but it still outperforms a single-tree approach. In this scenario, root nodes can be placed at the center of each side of the barbell. However, the backbone link connecting the clusters needs more resources and management than links in more evenly distributed topologies. Challenges also arise in topologies with long linear chains, posing difficulties for entanglement routing for all schemes.

## IV. SIMULATION RESULTS

This section presents our simulation settings and results.

### A. Simulation Settings

We have developed simulation models for multi-tree, single-tree, and synchronous routing schemes across three generated topologies: grid, random graph, and barbell. In all experiments, we maintain a constant direct-link entanglement generation success rate ($p$) and entanglement swapping success rate ($q$) at 0.8, along with a coherence time ($T_{co}$) of 2, which facilitates the comparability of results in various scenarios. Requests are started by choosing a source node at random and then selecting a destination node at random (across the entire network) from the group of nodes at the specified graph distance away.

The grid graph comprises a ten-by-ten network. In the case of multi-tree routing, we establish four root nodes positioned at the center of each quadrant, mirroring the configuration presented in Section III. Conversely, a single root node is designated at the center of the entire grid for single-tree routing.

A random topology graph is generated using the Erdős-Rényi model, also known as a binomial graph [12]. This model assigns probabilities for edge creation within the graph. Note that these probabilities only relate to the generation of physical topology. To ensure a node degree like the grid for comparison, we generate a random network with 100 nodes using a probability of 0.04. This yields an average of four adjacent edges per node.

The barbell topology comprises two clusters linked by a single connection. Each cluster constitutes a random graph containing 50 nodes, with an average of four edges per node. In the case of multi-tree routing, both clusters are further divided into two subgraphs, resulting in four root nodes. A single center is located at one end of the backbone link for single-tree routing.

The compared synchronous scheme relies on the two phases outlined in Section I and [8]. It operates on local knowledge of the instant topology. During the internal phase, each node assists in creating end-to-end entanglement for a user pair. Nodes keep a distance table and perform swapping operations for nodes closer to the source or destination. This synchronized approach involves exhaustive trials, utilizing all available entanglements.

### B. Results for Generated Topologies

Fig. 4 shows the entanglement rates for three routing schemes—multi-tree, single-tree structures, and synchronous (i.e., Syn) across three different network topologies—grid, random graph, and barbell over various distances. *Graph distance* is measured by the shortest path length between two nodes in the corresponding physical topology (i.e., hops). This article focuses on the impact of network topology rather than the actual distance and thus does not simulate $p$ as a function of distance. The mean entanglement rate is computed by dividing the number of successful end-to-end entanglements by the number of attempts.

The multi-tree approach shows advantages in the grid topology, characterized by its regular and evenly distributed layout. Conversely, a single-tree structure within a grid topology may lead to longer paths for nodes far from the central root node. In the random graph topology, nodes exhibit connectivity similar to the grid topology but are less evenly distributed. Therefore, they behave similarly to the grid topology but perform slightly worse. In the barbell topology, nodes within the same cluster enjoy better connectivity, but performance relies on the backbone link between clusters. The entanglement rate is substantially lower for pairs of nodes located on different sides of the barbell.

The "Syn" curve in Fig. 4, representing a synchronous routing scheme, typically exhibits lower rates than asynchronous multi-tree and single-tree schemes. This is because synchronous schemes may rely on coordination and blind searches, which may not scale well with distance, especially in less regular topologies. Overall, the improved performance of the multi-tree approach over the single-tree structure is attributable to node pairs in the same cluster, i.e., closer nodes.

*C. Simulations with Realistic Topologies*

To validate the effectiveness of the routing schemes in real-world scenarios, we also conducted simulations of the three routing methods on three actual research networks across the globe: ESnet [13], SURFnet [14], and Internet 2 [15]. These networks were explicitly selected to showcase the variance in graph structures, from a more uniformly structured SURFnet to a more linear configuration of Internet2. Note that distance in this context still refers to the shortest path length between two nodes in the physical topology rather than geographical distance.

SURFnet has nodes with strong connections, as indicated by their higher number of adjacent edges (e.g., Node 8 in Fig. 5a), making them suitable candidates for root nodes. This suggests that SURFnet exhibits areas of high network density, resulting in superior multi-tree performance compared to the other two topologies. As shown in Fig. 5b., ESnet exhibits less densely distributed nodes and fewer highly connected central nodes than SURFnet. Consequently, ESnet performs less efficiently than SURFnet but maintains similar relationships between the different routing approaches, as shown in Fig. 5b.

In the case of Internet 2, as shown in Fig. 5c, which includes long linear chains in its topology, DODAG behavior becomes less stable and inefficient. When two nodes are in the middle of a long linear chain, their likelihood of being included in the DODAG diminishes. Consequently, all three routing schemes exhibit suboptimal performance in this topology, with both multi-tree and single-tree schemes displaying overlapping performance. This instance is an example of a topology that lacks even connectivity and is not well-suited for multi-tree routing.

One potential solution to this is the inclusion of floating DODAGs. Floating DODAGs preserve detached branches as isolated DODAG segments when a direct-link entanglement is lost. During this period, the severed node temporarily serves as a provisional root for the subset until it successfully reconnects with the primary DODAG. Upon reconnection, the isolated segment reintegrates into the primary DODAG. It is important to note that only the provisional root node of a floating tree can join a stationary DODAG with a real root node. Further simulations involving floating trees are part of future work.

Overall, in realistic topologies, multi-tree routing shows a modest performance improvement compared to single-tree routing. This discrepancy can be attributed to the relatively modest sizes of the research networks, whereas the multi-tree approach is designed for large networks. In other words, the nodes in these realistic networks generally have shorter inter-node distances on average than those in the generated larger topologies.

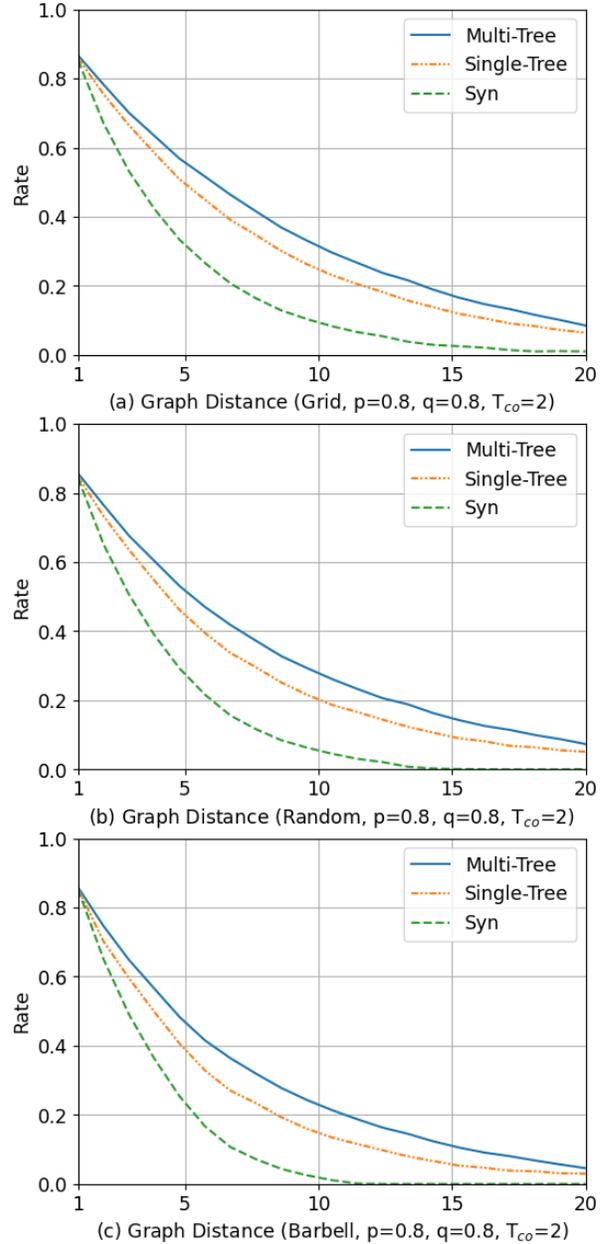

**Fig. 4.** Simulation results for generated topologies.

Nevertheless, the overarching trend suggests that asynchronous routing, in both single or multiple tree versions, outperforms synchronous routing in various simulated and realistic topologies.

V. CONCLUSION

Building upon the foundation laid by our previous work, which focused on asynchronous routing with local knowledge to improve end-to-end entanglement rates, this study proposes a multi-tree approach using a DODAG framework, addressing the challenge of long-distance entanglement distribution. Our simulations demonstrate improvements in entanglement rates across various network topologies, including grids and barbells, as well as real-world testbeds like ESnet and Internet2. These results highlight the efficiency and advantages of asynchronous routing schemes. This reinforces the potential of asynchronous strategies in optimizing future quantum networks.

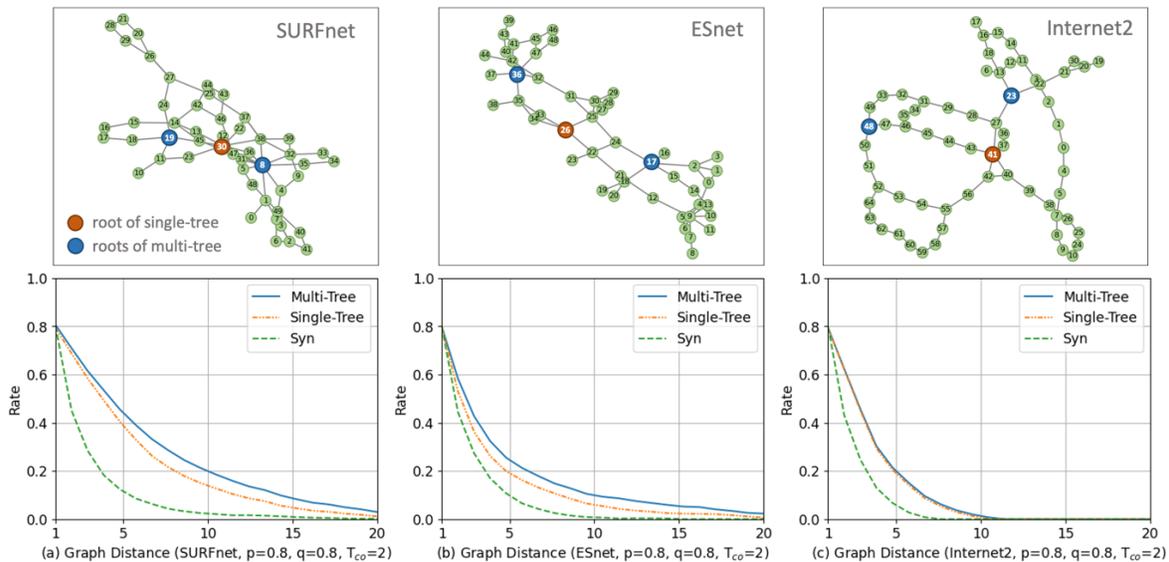

**Fig. 5.** Simulation results for realistic topologies: rate vs. distance for SURFnet, ESnet and Internet2.

ACKNOWLEDGMENT

This work was supported in part by Cisco University Research Gift #86944165, the Qatar Research, Development, and Innovation (QRDI) Academic Research Grant #ARG01-0501-230053, and in part by Prince Sattam Bin Abdulaziz University, Al-Kharj, Saudi Arabia. The findings achieved herein are solely the responsibility of the authors.

**Zebo Yang** (Graduate Student Member, IEEE) received an M.E. in computer engineering from Waseda University, Japan (2019). Currently pursuing a Ph.D. in computer science at Washington University in St. Louis, USA, he has previously worked as a Software Engineer at Tencent, Inc., Baidu, Inc., and DJI, Inc., from 2011 to 2017. Since 2019, he has been a Graduate Research Assistant at Washington University. His research interests include quantum computing, quantum networks, blockchains, and machine learning.

**Ali Ghubaish** (Graduate Student Member, IEEE) received a B.S. (Hons.) in computer engineering with a networking minor from Prince Sattam Bin Abdulaziz University, Saudi Arabia, in 2013. He obtained a M.S. in computer engineering from Washington University in St. Louis, USA, in 2017, where he is pursuing a Ph.D. Ali has been a Graduate Research Assistant at Washington University since 2018. His research focuses on network and system security, IoT, the Internet of Medical Things, and healthcare systems.

**Raj Jain** (Life Fellow, IEEE) received a B.E. in electrical engineering from APS University, India, and an M.E. from the Indian Institute of Science. He obtained his Ph.D. in Applied Maths (computer science) from Harvard University. He is the Barbara J. and Jerome R. Cox, Jr., Professor at Washington University in St. Louis. Dr. Jain co-founded Nayna Networks, Inc., and has held significant roles in academia and industry. He has received the 2018 James B. Eads Award and the 2017 ACM SIGCOMM Life-Time Achievement Award. He is known as one of the most cited authors in computer science and has authored "The Art of Computer Systems Performance Analysis." He is a Fellow of IEEE, ACM, and AAAS.

**Ramana Kompella** (Member, IEEE) received a B.Tech. from the Indian Institute of Technology, Bombay (1999), a M.S. from Stanford University (2001), and a Ph.D. from the University of California, San Diego (2007), all in computer science and engineering. He serves as a Distinguished Engineer and the Head of Research in Cisco's Emerging Tech and Incubation group, leading university research collaborations. His research focuses on data center networks, cloud performance optimization, and router algorithmics. Dr. Kompella has been an ACM member since 2007, receiving the NSF CAREER Award in 2011, and has received Best Paper awards at conferences like ACM SOCC.

**Hassan Shapourian** (Member, IEEE) received a M.S. in electrical engineering from Princeton University (2013) and a Ph.D. in Theoretical Physics from the University of Chicago (2019), and worked as a postdoctoral researcher at MIT/Harvard and Microsoft Station Q. He is a Senior Quantum Researcher at Cisco, where he leads projects on photonic quantum information processing and hardware physics. Dr. Shapourian is a recipient of Microsoft Research Postdoctoral Fellowship, Simons Postdoctoral Fellowship, Kavli Institute for Theoretical Physics (KITP) Graduate Fellowship, Biruni Graduate Student Research Award, John Bardeen Award (by UIUC), and Princeton University Graduate Fellowship.